\documentclass[superscriptaddress,twocolumn,showpacs,preprintnumbers,amsmath,amssymb]{revtex4}

\begin{document}

\preprint{SLAC-PUB-10841} 
\title{Fermionic sectors for the Kodama state}
\author{Stephon Alexander}
\email{stephon@itp.stanford.edu}
\affiliation{Department of Physics and SLAC, 
Stanford University,\\
2575 Sand Hill Road, Menlo Park CA, U.S.A \\ }
\author{Kristin Schleich}
\email{schleich@noether.physics.ubc.ca }
\author{ Donald 
M. Witt} 
\email{donwitt@noether.physics.ubc.ca }
\affiliation{Department of Physics and Astronomy, 
University of British Columbia,\\
6224, Agricultural Road, Vancouver, 
B.C., V6T 1Z1, Canada \\ }
\date{\today}

\begin{abstract}
{Diffeomorphisms not connected to the identity can act nontrivially on the quantum state space for gravity. However, in stark contrast to the case of nonabelian Yang-Mills field theories, for which the quantum state space is always in 1 dimensional representation of the large gauge transformations,  the quantum state space for gravity can have higher dimensional representations. In particular, the Kodama state will have 2 dimensional representations, that is sectors with spin 1/2, for many topologies that admit positive scalar curvature. The
existence of these spin 1/2 states are used to point out a possible answer to certain criticisms raised recently 
in the literature. }

\end{abstract}
\maketitle

It is well known that the state space for Yang-Mills field theory with nonabelian gauge groups in canonical quantization is invariant only under asymptotically trivial gauge transformations, i.e. those connected the identity of the gauge group. However,  non-abelian gauge groups also have gauge transformations not in the identity component, the so-called large gauge transformations. Equivalence classes of these large gauge transformations, equivalent under a gauge transformation in the identity component, form a group isomorphic to the group of additive integers \cite{Jackiw:1976pf,Callan:je}. Thus this group is Abelian and has only one-dimensional irreducible representations. Specifically, a Yang-Mills state vector $\Psi$ transforms under $g_n$, a large gauge transformation of degree $n$ as
\begin{equation}
g_n\Psi = e^{in\theta}\Psi\nonumber
\end{equation}
where the one dimensional representation is fixed by a single parameter, $\theta$. These are Yang-Mills $\theta$-states.

Although the appearance of $\theta$ states is most familiar for the case of Yang-Mills theories on spatial hypersurfaces with trivial topology, these groups (of equivalence classes of gauge transformations not in the identity component) are Abelian even for the case of nontrivial hypersurface topologies \cite{Isham:1981tq}. Hence $\theta$-states in Yang-Mills gauge theories only carry  one-dimensional irreducible representations of the large gauge transformations irrespective of hypersurface topology \footnote{In covariant quantization of the Yang-Mills theory, the $\theta$ term arises through the addition of a topological term, namely the first Chern class, to the action in the path integral. For nontrivial topology, the addition of the total derivative to the action will fail to produce the correct unitary irreducible representations. The correct answer must be obtained from the canonical approach.}.

In stark contrast, state vectors in gravity on spatial hypersurfaces of nontrivial topology can exhibit a much richer structure. In canonical quantum gravity, the diffeomorphism group plays the role of the nonabelian gauge group and, as in the case in Yang-Mills theories, the state space for quantum gravity is invariant only under diffeomorphisms connected to the identity. Correspondingly, the state space of quantum gravity carries the irreducible representations of equivalence classes of diffeomorphisms not in the identity component. Notably, in contrast to Yang-Mills field theories, the irreducible representations of these equivalence classes of diffeomorphisms not in the identity component need not be one-dimensional for all hypersurface topologies. In fact, these irreducible representations can carry spin 1/2 \cite{Friedman:1980st,Friedman:ft,Witt:ef}.

The goals of this paper 
are to provide a simple derivation of these spin 1/2 representations for an interesting class of 3-manifolds and to discuss their relevance to some recent comments about the Kodama state from  Witten \cite{Witten:2003mb}. 

The Kodama state \cite{Kodama:1990sc} is a holomorphic wavefunction expressed in terms of the Ashtekar connection. In the Ashtekar formulation of gravity, the geometrical content of the theory is expressed in terms of a  densitized inverse frame field $E^{a}{_i}$ indexed in a $SU(2)$ gauge group and a complex  left-handed $SU(2)$ connection $A_{a}^{i}$.  Thus the theory in these variables exhibits both diffeomorphism and gauge invariance. Precisely, the full invariance group of the theory is the semidirect product of the diffeomorphism group and the gauge group.

The Kodama state is given by the expression
\begin{equation}
 \Psi^{\pm}_{K}(A)={\cal N}{\rm e}^{\pm \frac{3 } { G \Lambda}\int Y_{CS}(A)}\label{Kstate}
\end{equation}
where the Chern Simons three form $Y_{CS}(A)$ is
\begin{equation}
Y_{CS}(A)=A\wedge dA + \frac{2}{3}A\wedge A\wedge A.
\end{equation}
The Ashtekar formulation is thus both gauge and diffeomorphism invariant and quantum states in this formulation are annihilated by both the gauge constraint and the Hamiltonian constraints of general relativity. This state is notable in that its annihilation by the constraints, for example
\begin {equation} \hat{{\cal H}}\rm^{grav}\Psi(A)_{K}=0 \end{equation}
\begin{equation}
{\cal H}^{grav} = \frac{1}{l_p} \epsilon_{ijk} E^{ai} E^{bj} (F_{ab}^k +
  G \Lambda\epsilon_{abc}E^{ck} )
\end{equation}
where $F_{ab}^k$ is the curvature of the complex $SU(2)$ connection $A_{ai}$, holds not simply in the semiclassical limit but as a quantum statement, that is with considerations of operator ordering. The semiclassical limit of this state is  a self-dual solution of these equations corresponding to de Sitter spacetime. Thus the Kodama state has been conjectured to be a candidate "ground state" for gravity with positive cosmological constant. This state has been considered mainly on $S^3$ topology. However, there are many 3-manifolds that admit semiclassical solutions with positive cosmological constant that are locally de Sitter spacetimes. Moreover, a special class of these solutions, 3-manifolds  $S^3/\Gamma$ where $\Gamma$ is a finite subgroup admit holomorphic wavefunctions of the minisuperspace form first studied by Kodama \cite{Kodama:1990sc}.

Why do the equivalence classes of the diffeomorphism group have a richer representation structure than those of nonabelian gauge theories? The answer lies in the fact that the elements of the diffeomorphism group are more restricted than those of a nonabelian gauge theory. A smooth diffeomorphism is a differentiable invertible map whose inverse is also differentiable. A gauge transformation is simply a map - its inverse is not required to exist. 

A convenient derivation of the $\theta$-states in Yang-Mills theories and their generalization for gravity  is that based on analyzing the topology of the reduced configuration space of the theory. When canonically quantizing a system whose configuration space is a manifold with nontrivial topology, it is well known that the quantum state vectors carry unitarily inequivalent irreducible representations of the fundamental group of the configuration space \cite{Balachandran:1991zj}. 

When quantizing theories with gauge or diffeomorphism invariance, the relevant configuration space is the reduced configuration space. For example, the reduced configuration space for Yang-Mills theory  in canonical quantization
 is ${\cal S}({\Sigma ^n})= {\cal A}({\Sigma ^n})/{\cal G}({\Sigma ^n})$ where ${\Sigma ^n}$ 
is the topology of the spatial hypersurface, ${\cal A}({\Sigma ^n})$ is the space of 
connections on ${\Sigma ^n}$ and ${\cal G}({\Sigma ^n})$ is the group of 
gauge transformations on ${\Sigma ^n}$. Similarly, the reduced configuration space  for gravity formulated in terms of a metric is
${\cal S}({\Sigma ^n})= {\rm Riem}({\Sigma ^n})/{\rm Diff}({\Sigma ^n})$ where ${\Sigma ^n}$ 
represents the spatial topology of our spacetime. ${\rm Riem}({\Sigma ^n})$ is the space of 
riemannian metrics on ${\Sigma ^n}$ and ${\rm Diff}({\Sigma ^n})$ is the group of 
diffeomorphisms of ${\Sigma ^n}$. The formulation of the reduced configuration space for gravity in terms of the Ashtekar variables is similar to that for gravity in terms of a metric, but complicated by the fact that the invariance is the semi-direct product of the diffeomorphism group and the gauge group. Furthermore, it is known that even in the case of trivial topology, the configuration space in Ashtekar variables under reduction  by only the gauge group ( gauge fixing the diffeomorphism group) has nontrivial fundamental group, leading to CP violation \cite{Ashtekar:1988sw}.  Under gauge fixing of the gauge group, the configuration space for Ashtekar variables again transforms under the diffeomorphism group and for sectors corresponding to nondegenerate frames, has the same structure as that of the space of riemannian metrics. So again the configuration space for Ashtekar variables will have a rich topological structure from the diffeomorphism group itself.

The reduced configuration space ${\cal S}({\Sigma ^n})$ is not a manifold unless the group  acts freely on the unreduced configuration space. For example, Yang-Mills 
connections with symmetries are fixed points in ${\cal A}({\Sigma ^n})$ and produce orbifold singularities in the reduced configuration space. However, these singularities 
can be removed either by removing the connections which admit symmetries from the space of connections or by restricting the gauge group to be a subgroup of ${\cal G}({\Sigma ^n})$ that does act freely, for example
${\cal G}_*({\Sigma ^n})\subset {\cal G}({\Sigma ^n})$, the group of all gauge 
transformations which fix a fiducial point in the space of connections. Similarly for gravity,  the configuration space ${\cal S}({\Sigma ^n})$ is not a 
manifold unless the group ${\rm Diff}({\Sigma ^n})$ acts freely on the unreduced configuration space. We denote the subgroup of the 
diffeomorphism group which acts freely as ${\rm Diff_A}({\Sigma ^n})$ as the group where ${\rm A}$ is 
condition on the diffeomorphism group such that the group action is free. Then
${\cal S}_{\rm A}({\Sigma ^n})= {\rm Riem}({\Sigma ^n})/{\rm Diff_A}({\Sigma ^n})$
is a manifold.

In the case of gravity, for a suitable choice of condition $\rm A$, one can always act on ${\cal S}_{\rm A}({\Sigma ^n})$ via a 
discrete group in order to obtain the ${\cal S}({\Sigma ^n})$. A natural choice of condition is to consider  ${\rm Diff_F}({\Sigma ^n})$ the group of all diffeomorphisms that leave a frame fixed at a point of the manifold. This subgroup of the diffeomorphism group acts freely on the configuration space as any isometry of the manifold necessarily changes either the basepoint or the frame\footnote{Note that the use of word isometry in this context refers to the topological structure of the manifold; isometries of the manifold need not be isometries of an arbitrary metric on the manifold. Of course,  there are metrics invariant under these isometries. For example, the isometry group of $S^3$ is $O^4$, the isometry group of its maximally symmetric metric, the round metric. Certain elements  act to map each point of the manifold into every other point, others, yield rotations around a fixed point. This example clearly illustrates that isometries do not leave a base point and frame invariant.}.  Now, in three dimensions, the isometry group of the manifold characterizes the homotopy structure of the diffeomorphism group according to the generalized Smale conjecture. This conjecture \cite{Hatcherconj}, precisely that the diffeomorphism group of a 3-manifold $\Sigma^3$ is contractible to its isometry group,  has been proven for many cases of the spherical spaces(Hatcher \cite{Hatcher} for $\Sigma^3=S^3$, the Smale conjecture itself ) . Hence frame fixing diffeomorphisms, which remove isometries act freely and ${\cal S}_{\rm F}({\Sigma ^n})$ is a manifold.

The fundamental group of the moduli space ${\cal S}_*({\Sigma ^n})$ for Yang-Mills theory  is easily calculated from the following exact sequence \cite{Steenrod}
\begin{eqnarray}
\dots \rightarrow& \pi _k({\cal G}_*({\Sigma ^n})) \rightarrow \pi _k({\cal A}({\Sigma ^n}))\rightarrow \pi _k({\cal S}_*({\Sigma ^n}))\rightarrow 
 \pi _{k-1}({\cal G}_*({\Sigma ^n}))\rightarrow \dots\cr &\dots\rightarrow \pi _0({\cal G}_*({\Sigma ^n})) 
\rightarrow \pi _0({\cal A}({\Sigma ^n}))\rightarrow \pi _0({\cal S}_*({\Sigma ^n}))
\label{exact}
\end{eqnarray}
The space ${\cal A}({\Sigma ^n})$ is always contractible to a point as the space of connections is affine\footnote{Given any two connections
$A_1$ and $A_2$, $\lambda A_1 + (1- \lambda )A_2$ is connection for all $\lambda $ with $0\le \lambda
\le 1$. } Hence $\pi _k({\cal A}({\Sigma ^n}))=1$ and the exact sequence yields 
$$1\rightarrow \pi _k({\cal S}_*({\Sigma ^n}))\rightarrow \pi _{k-1}({\cal G}_*({\Sigma ^n}))\rightarrow 1.$$
This implies that $\pi _k({\cal S}_*({\Sigma ^n}))= \pi _{k-1}({\cal G}_*({\Sigma ^n}))$. The homotopy groups
of ${\cal G}_*({\Sigma ^n})$ are just the homotopy groups of the space of functions from ${\Sigma ^n}$ to
$G$. As ${\cal G}_*({\Sigma ^n})$ is a group, all of its homotopy groups except $\pi _0$ are abelian. 

The group
$\pi _0({\cal G}_*({\Sigma ^n}))$ is the 0-th homotopy class of maps from ${\Sigma ^n}$ to $G$ and in all cases can be explicitly calculated from the cohomology using obstruction theory. The abelian nature of the cohomology yields the the fact that  $\pi _1({\cal S}_*({\Sigma ^n}))= \pi _{0}({\cal G}_*({\Sigma ^n}))$ is abelian. 
This is immediately apparent in the special case
where ${\Sigma ^n}=S^n$ for which $\pi _0({\cal G}_*({S^n}))=\pi _n(G)$ as the homotopy class of maps from $S^n$ into $G$ is by definition $\pi _n(G)$. Thus the fundamental group of ${\cal S}_*({\Sigma ^n})$ is always abelian and the only irreducible 
representations of it are consequently one dimensional. This means that no 2 dimensional representations, i.e. spinorial representations, can occur for
Yang-Mills. In fact when ${\Sigma ^n}$ is an n-sphere and the group $G$ is semi-simple
the fundamental group always contains a copy of the integers. This implies that  for this case, there is always a continuous parameter in the theory $\theta$ which can be tuned. Thus this parameter can be tuned to small values in order to set physical effects arising from its presence to agree with experimental data.

The fundamental group of the moduli space ${\cal S}_F({\Sigma ^n})$ for gravity is also calculated from the exact sequence (\ref{exact}).
The total space for gravity is the space ${\rm Riem}({\Sigma ^n})$ is contractible so again
the exact sequence (\ref{exact}) implies that 
\begin {equation}
\pi _1{\cal S}_{\rm F}({\Sigma ^n})=\pi _0{\rm Diff_F}({\Sigma ^n}).
\end{equation}
However, in sharp contrast to the case of Yang-Mills theory, the group $\pi _0{\rm Diff_F}({\Sigma ^n})$ cannot be calculated
from cohomology and obstruction theory. The reason is unique to the properties of the diffeomorphism group.  Diffeomorphisms  are by definition
differentiable maps with differentiable inverses, in contrast to gauge transformations of Yang-Mills theory.

For $n=3$, the 
fundamental group of ${\cal S}_{\rm F}({\Sigma ^n})$ is typically non-trivial, implying that
the quantum  state space of gravity is defined on a multiply connected configuration space. The 
quantum states carry unitary irreducible representations of the group 
$\pi _1{\cal S}_{\rm F}({\Sigma ^n})$. The generic situation is that for most ${\Sigma ^3}$ there are spinorial representations.

Of particular interest to the Kodama state is the case of the spherical spaces,  $\Sigma ^3=S^3/\Gamma$, where $\Gamma$ is a finite group that freely acts on $S^3$.  In this case, the fundamental groups of ${\cal S}_{\rm F}({\Sigma ^3})$ are $SU(2)$ coverings of non-Abelian $SO(3)$ crystal groups \cite{Witt:ef}.  A simple example of a space ${\Sigma }^3$ with rich non-Abelian structure for is the quaternionic space 
$S^3/D^*_8$. $D^*_8$ is the group of quaternions, with elements 
$\{\pm 1, \pm i,\pm j,\pm k \}$. As $S^3 $ is the group manifold $SU(2)$,  a nice representation of $D^*_8$ is given by the pauli matrices and the space $S^3/D^*_8$  is easily realized as a coset space. A computation \cite{Witt:ef} yields $\pi_0\hbox{\rm Diff}_{F}(S^3/D^*_8)= O^*$, the binary octohedral group \footnote{The binary octohedral group has a presentation $< x,y: x^2 = (xy)^3 = y^4, x^4 = 1>$.}. In contrast, the computation for orientation preserving diffeomorphisms ${\rm Diff}^+$
 yields $\pi_0\hbox{\rm Diff}^{+}(S^3/D^*_8)= O$ the octohedral group \footnote{The octohedral group has presentation $< x,y: x^2 = (xy)^3 = y^4, x^4 = 1>$.}.

 A $2\pi$ rotation in ${\rm Diff}_{F}(\Sigma^3)$ can be constructed as follows:  take $B^3$  and $B'^3$ to be two open concentric balls in $\Sigma^3$ centered around the point with fixed frame. Let $f$ be a diffeomorphism of $\Sigma$ which is the identity in  $B^3$ and in the complement of $B'^3$ in $\Sigma^3$ and corresponds to a $2\pi$ rotation in $B^3\cap B'^3$.
Intuitively this diffeomorphism corresponds to rotating the ball $B^3$ by $2\pi$ while keeping all but a shell of its complement in the manifold $\Sigma^3$ also fixed.

A diffeomorphism  implementing a $2\pi$ rotation  is deformable to the identity in $\pi_0\hbox{\rm Diff}^{+}(S^3/D^*_8)= O$, but not in $\pi_0\hbox{\rm Diff}_{F}(S^3/D^*_8)= O^*$. This can be understood intuitively as follows: In the case of $\pi_0\hbox{\rm Diff}^{+}$, deformations can be extended into the ball $B^3$. In particular, a deformation  corresponding to allowing the ball to counterrotate about a fixed point to undo the $2\pi$ rotation in the shell $B^3\cap B'^3$ is allowed. Hence the $2\pi$ rotations is deformable to the identity in $\pi_0\hbox{\rm Diff}^{+}(S^3/D^*_8)$. However, such a deformation cannot be carried out in $\pi_0\hbox{\rm Diff}_{F}(S^3/D^*_8)= O^*$; any deformation that results in a rotation about a fixed point would rotate the frame and hence is not an element. Moreover, there is no deformation to the identity with support in the  complement of $B'^3$ in $\Sigma^3$ because $\pi_0\hbox{\rm Diff}_{F}(S^3/D^*_8)= O^*$ is a subgroup of $SU(2)$ but not of $SO(3)$.  However a diffeomorphism implementing a $4\pi$ rotation is deformable to the identity in  $\pi_0\hbox{\rm Diff}_{F}(S^3/D^*_8)= O^*$. Hence the zeroth homotopy group of frame fixing diffeomorphisms of $S^3/D^*_8$ admits a spin 1/2 representation, precisely a 2-dimensional representation that transforms under rotations via a unitary representation of $SU(2)$ indexed by $\pi_0\hbox{\rm Diff}_{F}(S^3/D^*_8)$.

Indeed, the spherical spaces  $\Sigma ^3=S^3/\Gamma$ for $\Gamma$ a non-cyclic group always admit such a spin 1/2 representation  (see table IV of \cite{Witt:ef} and in fact only admit 1-d and 2-d representations. Furthermore, the connected sum of two closed 3-manifolds that admit positive scalar curvature and a 2-d representation  will also admit  both 1-d and 2-d representations.

Note that in this case there is 
no free parameter as in the case of traditional $\theta  $-states in Yang-Mills.  One can choose the representation, but having done so, its structure is fixed. Thus the specification of the representation is an essential part of the initial conditions for the quantum state space and once chosen, cannot be tuned by a physical mechanism.

These $SU(2)$ coverings of non-Abelian $SO(3)$ crystal groups
groups also occur in the finite theory of quantum shapes \cite{Balachandran:1991ea}. As in the above case, these quantum states for quantum shapes  have no $\theta $ parameter which can be tuned. The irreducible unitary representations indexed by the fundamental group of the configuration space  are either present or not present. Thus as in the gravitational case, the theory of quantum shapes admits fermionic representations.

What are the implications of the existence of spinorial representations in gravity for the Kodama state? The Kodama wavefunction is explicitly  written in (\ref{Kstate}) as a 1 dimensional representation of  
$\pi _0{\rm Diff_F}(\Sigma^3)$. This is apparent as it is the exponential of an action, or a scalar function of the geometry. Hence, as the action is invariant under all diffeomorphisms, the state will be invariant under the action of any diffeomorphism corresponding to a $2\pi$ rotation. However, we can construct a generalized Kodama state which is in the 2 dimensional representation. Let $f$ be a frame fixing diffeomorphism on a 3-manifold $\Sigma^3$ that has a 2 dimensional representation. Then

\begin{equation}\Phi_K (A)= \sum_{[f]\in {\pi _0}\hbox{\rm Diff}_F(\Sigma^3)} \chi([f]) \Psi^{\pm}_{K}(A)\label{spinK}\end{equation} 
where $[f] \in {\pi _0}\hbox{\rm Diff}_F(\Sigma^3)$ is a sum over equivalence classes of frame fixing diffeomorphisms,  $\chi([f])$ its character and $\Psi^{\pm}_{K}(A) $ is  given by (\ref{Kstate}). $\Phi_K$ will transform as a spin 1/2 representation. Moreover, the constraints will be exactly satisfied by this wavefunction by construction. 

Furthermore, there is no orientation reversing diffeomorphism $P$ for $S^3/\Gamma$ that admit spin 1/2 states \cite{Witt:ef}. This property holds for all representations, not just the 2 dimensional ones but is correlated with the appearance of spin 1/2 states.

These properties of a spinorial Kodama state could potentially change the conclusions of Witten for such 3-manifolds. Witten studied the normalizability of the Kodama state for the Abelian and Yang-Mills gauge theories and found that the Hamiltonian constraint admitted both a normalizable and non normalizable part.  This can be explicitly be seen, for example, by noting that  if one chooses a normalizable Kodama state and acts on it with CP , the state will transform into a unnormalizable Kodama state.  This happens because the Chern-Simons functional is CP odd.  A Fock space can be constructed from expanding about the Kodama state and one finds that gravitons of one helicity have positive energy and those of the opposite helicity have negative energy. 

For the spinorial Kodama state, as discussed above, there is no definition of parity. Hence, what is meant by CP for such manifolds is not clear. In particular there will be no map which can reverse the sign of the Chern-Simons action and thus interchange Chern-Simons wavefunctions.  In addition, for such spinorial states, one anticipates that the association of the helicity of gravitons with the positivity or negativity of their energy will fail, as it may not be possible to associate a global helicity with linearized gravitons. Finally, it is natural to ask if the Kodama state itself is spinorial, are the linearized fluctuations in some way also spinorial? 

The fact that spinorial states exist for the Kodama state is not the end of the story. One must perform a more careful analysis of the properties of these states to determine the properties of these fermionic states and fluctuations around these fermionic states. Such work is in progress.

A.S. was supported by US DOE under grant DE-AC03-76SF00515. K.S.  and D.W. were supported in part
by the Natural and Sciences and Research Council of Canada. In addition, K.S. and D.W. would like
thank the Perimeter Institute for Theoretical Physics for their hospitality during the completion of part of this work.

\end{document}